\documentclass[aps]{revtex4}
\usepackage{amssymb,epsf}

\begin{document}

\title{Black hole solutions correspondence between conformal and massive
theories of gravity}
\author{B. Eslam Panah $^{1,2}$\footnote{
email address: beslampanah@shirazu.ac.ir}, and S. H. Hendi$^{1}$\footnote{%
email address: hendi@shirazu.ac.ir}}
\affiliation{$^1$ Physics Department and Biruni Observatory, College of Sciences, Shiraz
University, Shiraz 71454, Iran\\
$^2$ Research Institute for Astronomy and Astrophysics of Maragha (RIAAM),
Maragha, Iran}

\begin{abstract}
In this work, a correspondence between black hole solutions of
conformal and massive theories of gravity is found. It is seen
that this correspondence imposes some constraints on parameters of
these theories. What is more, a relation between the mass of black
holes and the parameters of massive gravity is found. Indeed, the
acceptable ranges of massive gravity parameters ($c_{1}$\ and
$c_{2}$) are found. It is shown that by considering the positive
mass of black holes, some ranges of $c_{1}$\ and $c_{2}$ are
acceptable.
\end{abstract}

\maketitle

\section{Introduction}

The general relativity (GR) is a successful theory for explaining
the effects of gravity within the solar system such as the
precession of the perihelion of Mercury and the gravitational
bending of light around the Sun. Nevertheless, there are some
puzzles left on scales beyond the distances of the solar system.
For example, the observational evidence of galactic rotation
curves is not consistent with the predictions of GR: the unknown
dark matter. In addition, GR cannot describe the accelerated
expansion of the Universe. In order to provide the energy source
to explain this acceleration, the concept of dark energy is
introduced. According to the mentioned reasons, the modification
of GR is necessary. One of the modified theories of gravity is the
conformal (Weyl) gravity, whose action is defined by the square of
the Weyl tensor
\cite{EnglertTG,Narlikar,Riegert,Maldacena,LuP,Bars}. This theory
is invariant under a conformal transformation of the metric
tensor as%
\begin{equation}
g_{\mu \nu }\rightarrow g_{\mu \nu }^{\ast }=\Omega ^{2}g_{\mu \nu },
\end{equation}%
where $\Omega =\Omega \left( x\right) $ is a nonsingular function of
spacetime coordinates.

It is notable that, the solutions of the GR equation are a subset
of the solutions of conformal gravity. It was shown that conformal
gravity with a Neumann boundary condition can select the GR
solution out of conformal gravity \cite{Maldacena,AnastasiouO}.
According to the fact that the conformal gravity possesses more
solutions than GR, and also it can explain
dark matter and dark energy scenarios \cite%
{MannheimK,Mannheim2006,MannheimO,Mannheim2012}, this motivates us
to consider the conformal gravity theory in this work.

Strictly speaking, conformal theory of gravity has some
interesting properties, e.g., \ i) it is useful for constructing
super-gravity theories \cite{BergshoeffRd,deWit}.\ ii) it can be
considered as a possible UV completion of GR theory
\cite{Adler,Hooft,Mannheim2012}.\textbf{\ }iii) it
can solve the problem of spacetime singularities \cite%
{BambiMR,ChakrabartyBBM}. iv) it was shown that the observational data
(X-ray data) of astrophysical black holes is compatible with this theory
\cite{Zhouetal}.\textbf{\ }v) it also arises from twister-string theory with
both gauge-singlet open strings and closed strings \cite{BerkovitsW}.\textbf{%
\ }vi) this theory can explain the dark side of the Universe \cite%
{MannheimK,Mannheim2006,MannheimO,Mannheim2012}. vii) it can
appear as a counterterm in AdS/CFT\ calculations
\cite{LiuT,Balasubramanian}.

Another modification of GR is represented by massive gravity
theories. From the perspective of modern particle physics,
graviton is a massless spin-2 particle in GR. In order to endow
mass to the graviton, massive gravity was introduced by Fierz and
Pauli in 1939 \cite{FP}. It is notable that this theory included
massive graviton in flat space (linear level) where in non-flat
background\ (non-linear level) it encountered the Boulware-Deser
ghost \cite{BD}. Fortunately, a version of the ghost-free massive
theory was proposed by de Rham,
Gabadadze and Tolley which is known as dGRT massive gravity \cite%
{dRG,dRGT,dRGTI,dRGTII}. In ref. \cite{Kareeso}, the mass-radius ratio
bounds for compact objects have been studied in this gravity. Also, Yamazaki
et al. \cite{Yamazaki}, discussed the boundary conditions for the
relativistic stars and obtained the mass-radius relation of stars in this
gravity. Black hole solutions and their thermodynamical quantities in the
dRGT theory were investigated in refs. \cite%
{BHMassI,BHMassII,BHMassIII,BHMassIV,BHMassV,BHMassVI,BHMassVII,BHMassVIII,BHMassIX,BHMassX,BHMassXI}%
.

It is notable that, modification of reference metric related to
the physical metric of spacetime leads to the possibility of
introduction of different classes of dRGT-like massive gravity
\cite{dRGTII}. One of the massive theories of gravity is proposed
by Vegh \cite{Vegh}, which is applied in gauge/gravity duality. In
other words, this theory is similar to the dRGT massive gravity
with a difference in its reference metric which is a singular one.
Graviton in Vegh's massive gravity may behave like a lattice and
exhibits a Drude peak \cite{Vegh}. In addition, it was shown that
for an arbitrary singular metric, this theory of massive gravity
is ghost free and stable \cite{ZhangL}.

Generally, massive gravity has some interesting properties e.g.,
i) similar to conformal gravity, this theory of gravity would
provide a solution for the cosmological constant problem
\cite{DvaliGS,DvaliHK} and also can explain the self-acceleration
of the Universe without adding the cosmological constant to the
field equation \cite{DeffayetDG}. In other words, one of the
massive gravity terms plays the role of cosmological constant
\cite{GumrukcuogluLM,Gratia}. ii) it results into extra
polarization for gravitational waves, and affects the
propagation's speed of gravitational waves \cite{Will}, and also
producs the gravitational waves during inflation
\cite{Mohseni,GumrukcuogluI}. iii) the existence
super-Chandrasekhar white dwarfs with mass greater that 1.45 times
the mass of the Sun \cite{white}, and also of massive neutron
stars with mass greater than three times that of the Sun
\cite{Neutron}. iv) the topological black holes in this theory can
exhibit van der Waals behavior \cite{vanmass}, and heat engines
\cite{heat}.\ v) the existence of a black hole remnant in massive
gravity with a possible candidate for dark matter \cite{remnant}.

Taking into account some similar results of conformal gravity and
massive theory, one may be motivated to find a possible
fundamental relation between them. In this paper we want to show
that there is a correspondence between the black hole solutions in
conformal gravity and those of Vegh's massive gravity. In other
words, we will show that the exact solutions of the metric
function in conformal gravity can be the same, as the
corresponding solutions of Vegh's massive gravity in a special
case. We also use this correspondence to obtain some constraints
on the massive parameters.

\section{Black Hole Solutions in Conformal Gravity}

The $4$-dimensional action of conformal gravity is given by a
square of the Weyl tensor in the following form:
\begin{equation}
I_{G}=-\alpha \int d^{4}x\sqrt{-g}C^{\mu \nu \rho \sigma }C_{\mu \nu \rho
\sigma },  \label{eq16}
\end{equation}%
where $\alpha $ is a dimensionless coupling constant and plays an important
role in critical gravity \cite{Bergshoeff,Alishahiha,Gullu}. Hereafter, we
set $\alpha =1$ without loss of generality.

Varying the action (\ref{eq16}), with respect to the metric tensor
$g_{\mu \nu }$, leads to the following equation:
\begin{equation}
\left( 2\nabla ^{\rho }\nabla ^{\sigma }+R^{\rho \sigma }\right) C_{\mu \rho
\sigma \nu }=0.  \label{fiedEq1}
\end{equation}

Now we consider the field eq. (\ref{fiedEq1}), and a
$4$-dimensional general spacetime in the following form:
\begin{equation}
ds^{2}=-f(r)dt^{2}+f^{-1}(r)dr^{2}+r^{2}h_{ij}dx_{i}dx_{j},~~i,j=1,2,
\label{Metric}
\end{equation}%
in which $h_{ij}dx_{i}dx_{j}$ is a $2$-dimension line-element for an
Euclidian space with constant curvature $2k$ and volume $V_{2}$, as
\begin{equation}
h_{ij}dx_{i}dx_{j}=\left\{
\begin{array}{cc}
d\theta ^{2}+d\varphi ^{2} & k=0 \\
d\theta ^{2}+\sin ^{2}\theta d\varphi ^{2} & k=1 \\
d\theta ^{2}+\sinh ^{2}\theta d\varphi ^{2} & k=-1%
\end{array}%
\right. .
\end{equation}

The solution of conformal gravity has been extracted in refs. \cite%
{Klemm,Mannheim1991} as\textbf{\ }%
\begin{equation}
f\left( r\right) =s_{0}+s_{1}r+\frac{s_{2}}{r}-\frac{\Lambda }{3}r^{2},
\label{F(r)con}
\end{equation}%
where $s_{0}$, $s_{1}$, $s_{2}$ and $\Lambda$ are four constants.
Due to the fact that the nonzero components of the field equation
are, at least, third-order differential equations, one expects
three integration constants. In order to extract the solution
(\ref{F(r)con}), we have to restrict three of these constants
($s_{0}$, $s_{1}$, and $s_{2}$) in the following constraint, as
(see ref. \cite{Mannheim1991}, for more details)
\begin{equation}
s_{0}^{2}=3s_{1}s_{2}+k^{2}.  \label{cons}
\end{equation}

Considering $s_{1}=0$, this solution reduces to Schwarzschild
(A)dS spacetime. Another interesting constant of the solution
(\ref{F(r)con}), is called $\Lambda$. This constant plays the role
of cosmological constant. It is notable that $\Lambda$ as an
integral constant and has not been inserted in the action by hand.
Indeed, we can extract the cosmological constant by solving the
field equation of conformal gravity (\ref{fiedEq1}), by using the
mentioned metric (\ref{Metric}). In other words, conformal gravity
would solve the dark matter problem and explains the current
acceleration of our Universe.

Now, we briefly examine the geometrical structure of this
solution. Our aim is to establish the possibility of having black
hole solution. For this goal, we first look for the essential
singularity(ies) by investigating two
scalar curvatures: Ricci and Kretschmann scalars. Considering the metric (%
\ref{Metric}), with the (\ref{F(r)con}), the Ricci scalar is extracted as
\begin{equation}
R=4\Lambda -\frac{6s_{1}}{r}+\frac{2\left( 1-s_{0}\right) }{r^{2}}
\end{equation}

Evidently, the Ricci scalar diverges at the origin ($\lim_{r\longrightarrow
0}R\longrightarrow \infty $). On the other hand, the Kretschmann scalar is
given by
\begin{eqnarray}
K &=&\frac{8\Lambda ^{2}}{3}-\frac{8\Lambda s_{1}}{r}+\frac{8\left(
s_{1}^{2}+\frac{\Lambda }{3}\left( 1-s_{0}\right) \right) }{r^{2}}-\frac{%
8s_{1}\left( 1-s_{0}\right) }{r^{3}}  \nonumber \\
&&  \nonumber \\
&&+\frac{4\left( 1-s_{0}\right) ^{2}}{r^{4}}-\frac{8s_{2}\left(
1-s_{0}\right) }{r^{5}}+\frac{12s_{2}^{2}}{r^{6}},
\end{eqnarray}%
\bigskip

\begin{figure}[t]
\centering
$%
\begin{array}{c}
\epsfxsize=8cm \epsffile{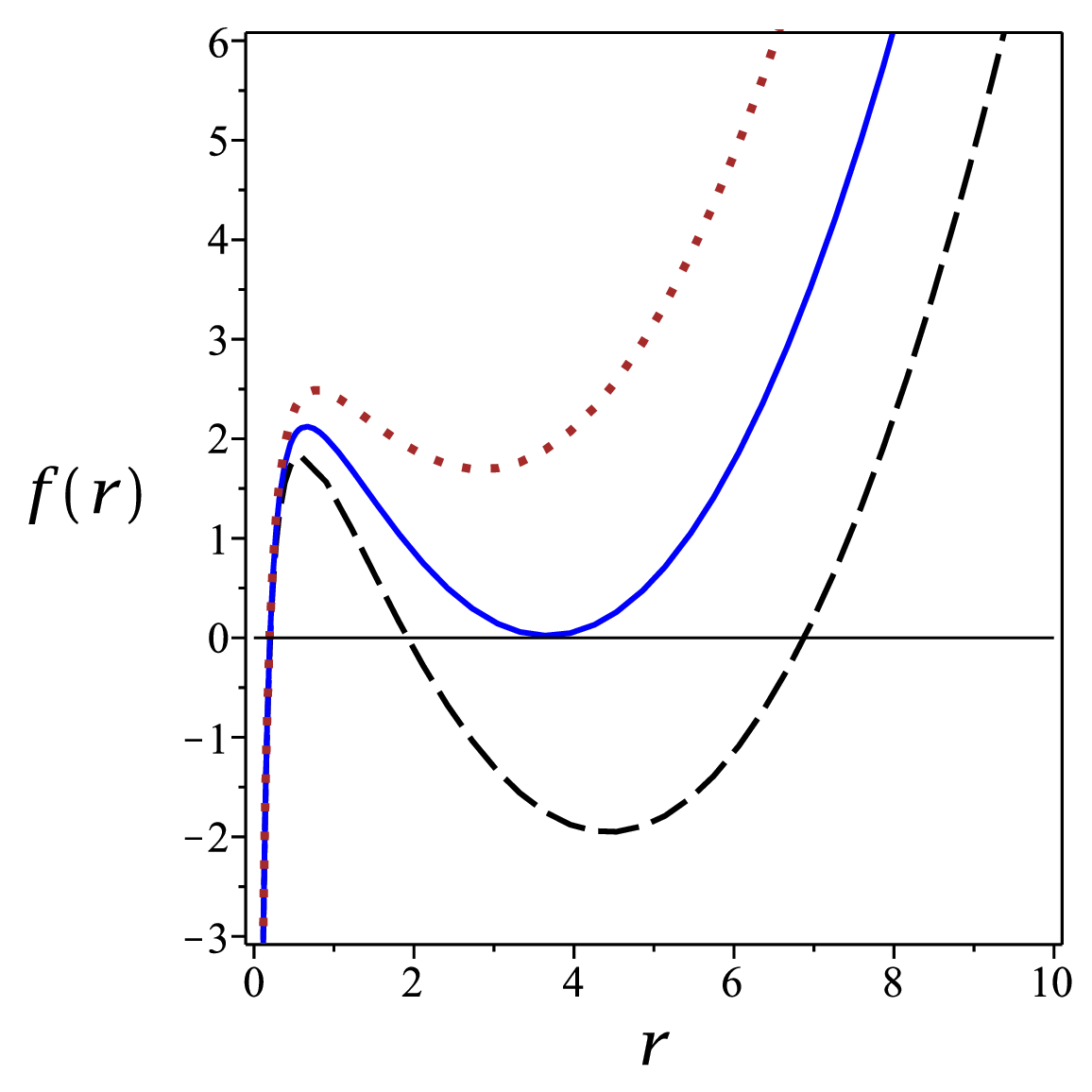}%
\end{array}
$%
\caption{$f(r)$ versus $r$ for $\Lambda=-1$, $s_{0}=5.0$, $s_{2}=-0.9$, $%
s_{1}=-3.0$ (dashed line), $s_{1}=-2.51$ (continuous line) and $s_{1}=-2.0$
(dotted line).}
\label{Fig1}
\end{figure}

One can show that this scalar has the following behavior:
\begin{eqnarray}
\lim_{r\longrightarrow 0}R_{\alpha \beta \gamma \delta }R^{\alpha \beta
\gamma \delta } &\longrightarrow &\infty , \\
&&  \nonumber \\
\lim_{r\longrightarrow \infty }R_{\alpha \beta \gamma \delta }R^{\alpha
\beta \gamma \delta } &\longrightarrow &\frac{8\Lambda ^{2}}{3},
\end{eqnarray}%
which can confirm that: i) there is a curvature singularity at
$r=0$. ii) the asymptotic behavior of this solution is (A)dS,
since the Kretschmann scalar is $\frac{8\Lambda ^{2}}{3}$ at
$r\longrightarrow \infty $. Also we plot the solution
(\ref{F(r)con}) versus $r$ in Fig. \ref{Fig1}, which shows that it
can be covered by an event horizon. In other words, this solution
would be interpreted as a black hole. Another important property
of black holes in conformal gravity is related to the existence of
multi-horizons similar to black holes in massive gravity
\cite{multiH} (see Fig. \ref{Fig1} for more details). Indeed, it
is possible to find more than two horizons for black holes in
conformal gravity. Thermodynamic properties of this solution have
been studied in some literatures. \cite{LiuPJ,XuZhao,XuSZ}.

\section{Black Hole Solutions in Massive Gravity}

The action of the Einstein gravity by adding massive terms and in
the presence the cosmological constant ($\Lambda $) is given by
\cite{Vegh}
\begin{equation}
I=\frac{1}{2\kappa ^{2}}\int d^{4}x\sqrt{-g}\left[ R-2\Lambda
+m^{2}\sum_{i}^{4}c_{i}\mathcal{U}_{i}\left( g,f\right) \right] ,
\label{Action}
\end{equation}%
in the above action $R$ and $m$ are the Ricci scalar and the mass
of the graviton, respectively, and $\Lambda =\frac{-3}{l^{2}}$.
Also, $g$ and $f$ are a metric tensor and a fixed symmetric
tensor, respectively. It is notable that $c_{i}$'s are arbitrary
constants whose values can be determined according to
observational or theoretical considerations. In addition,
$U_{i}$'s are symmetric polynomials of the eigenvalues of the
$d\times d$ matrix $K_{\nu }^{\mu }=\sqrt{g^{\mu \alpha }f_{\alpha
\nu }}$. Indeed, $U_{i}$'s are interaction terms which introduced
in the following forms \cite{Vegh}:
\begin{eqnarray}
\mathcal{U}_{1} &=&\left[ K\right] ,~~~~~\mathcal{U}_{2}=\left[ K\right]
^{2}-\left[ K^{2}\right] ,~~~~~\mathcal{U}_{3}=\left[ K\right] ^{3}-3\left[ K%
\right] \left[ K^{2}\right] +2\left[ K^{3}\right] ,  \nonumber \\
\mathcal{U}_{4} &=&\left[ K\right] ^{4}-6\left[ K^{2}\right] \left[ K\right]
^{2}+8\left[ K^{3}\right] \left[ K\right] +3\left[ K^{2}\right] ^{2}-6\left[
K^{4}\right] ,  \nonumber \\
\mathcal{U}_{5} &=&\left[ K\right] ^{5}-10\left[ K\right] ^{3}\left[ K^{2}%
\right] +20\left[ K\right] ^{2}\left[ K^{3}\right] -20\left[ K^{2}\right] %
\left[ K^{3}\right] +15\left[ K\right] \left[ K^{2}\right] ^{2}  \nonumber \\
&&-30\left[ K\right] \left[ K^{4}\right] +24\left[ K^{5}\right] ,  \nonumber
\\
&&...~.
\end{eqnarray}

Using the action (\ref{Action}) and varying it with respect to the metric
tensor, $g_{\mu \nu }$, one obtains
\begin{equation}
G_{\mu \nu }+\Lambda g_{\mu \nu }+m^{2}X_{\mu \nu }=0,  \label{feq}
\end{equation}%
where $G_{\mu \nu }$\ is Einstein's tensor. In the field equation (\ref%
{feq}), $X_{\mu \nu }$\ is the massive term which is given by
\begin{eqnarray}
X_{\mu \nu } &=&-\frac{c_{1}}{2}\left( \mathcal{U}_{1}g_{\mu \nu }-K_{\mu
\nu }\right) -\frac{c_{2}}{2}\left( \mathcal{U}_{2}g_{\mu \nu }-2\mathcal{U}%
_{1}K_{\mu \nu }+2K_{\mu \nu }^{2}\right)   \nonumber \\
&&  \nonumber \\
&&-\frac{c_{3}}{2}\left( \mathcal{U}_{3}g_{\mu \nu }-3\mathcal{U}_{2}K_{\mu
\nu }+6\mathcal{U}_{1}K_{\mu \nu }^{2}-6K_{\mu \nu }^{3}\right) -\frac{c_{4}%
}{2}\left( \mathcal{U}_{4}g_{\mu \nu }-4\mathcal{U}_{3}K_{\mu \nu }+12%
\mathcal{U}_{2}K_{\mu \nu }^{2}\right.   \nonumber \\
&&  \nonumber \\
&&\left. -24\mathcal{U}_{1}K_{\mu \nu }^{3}+24K_{\mu \nu }^{4}\right) +...~.
\end{eqnarray}

In order to extract the static charged black holes in the context
of massive gravity with (A)dS asymptote, we consider the metric
(\ref{Metric}), and an appropriate reference metric introduced in
refs. \cite{Vegh,Cai2015,HendiEP},
\begin{equation}
f_{\mu \nu }=diag(0,0,\mathcal{C}^{2}h_{ij}),  \label{f11}
\end{equation}%
where $\mathcal{C}$\ is a positive constant. Considering the above reference
metric, for $4$-dimensional spacetime, $U_{i}$'s are as \cite%
{Cai2015,HendiEP}
\begin{eqnarray}
\mathcal{U}_{1} &=&\frac{2\mathcal{C}}{r},  \nonumber \\
\mathcal{U}_{2} &=&\frac{2\mathcal{C}^{2}}{r^{2}},  \nonumber \\
\mathcal{U}_{i} &=&0,~~~~i>2.
\end{eqnarray}

Using the metric (\ref{Metric}) with the field equation (\ref{feq}), the
metric function $f(r)$\ is obtained in the following form \cite%
{Cai2015,HendiEP}:
\begin{equation}
f\left( r\right) =k-\frac{M}{r}-\frac{\Lambda }{3}r^{2}+m^{2}\left( \frac{%
\mathcal{C}c_{1}}{2}r+\mathcal{C}^{2}c_{2}\right) ,  \label{f(r)}
\end{equation}%
where $M$ is an integration constant related to the total mass of black
holes. It is shown that the solutions (\ref{f(r)}) can be interpreted as
black holes with more than two horizons \cite{multiH,Cai2015,HendiEP}.
The thermodynamic aspects of this solution have been studied in \cite%
{BHMassIII,vanmass}.

\section{The correspondence between black hole solutions of conformal and
massive gravities}

Here, we want to show that the black hole solutions in conformal
and massive theories of gravity are the same. In other words, we
indicate that these theories of gravity describe the same black
holes. For this goal, we compare these solutions. Our analysis
shows that these solutions are the same provided we impose some
constraints on parameters of conformal and massive gravities in
the following forms:
\begin{eqnarray}
s_{0} &=&k+m^{2}\mathcal{C}^{2}c_{2},  \nonumber \\
s_{1} &=&\frac{m^{2}\mathcal{C}c_{1}}{2},  \nonumber \\
s_{2} &=&-M.  \label{para}
\end{eqnarray}

Adjusting the parameters of conformal and massive gravities in the
above equation, black hole solutions in these theories are the
same. Here, we can ask why these solutions are the same. Although
the answer to this question is not trivial, some similar
properties help us to deepen our insight for these theories. For
example; i) both of these theories can explain the dark matter.
ii) both of them nicely address the cosmological constant problem
and explain the self-acceleration of the Universe without
introducing the cosmological constant. iii) these gravities
consider the quantum nature of black holes. iv) both of black
holes have multi horizons.

\subsection{Physical limitations based on this correspondence}

Another interesting result of this correspondence is related to
the mass of black holes in massive gravity. Indeed, we can extract
a relation for the mass of black holes which depends on the
parameters of massive gravity. Considering the adjustment
(\ref{para}) and putting it into eq. (\ref{cons}), we obtain a
relation for the mass of black hole versus the parameters of
massive gravity and the topological factor ($k$) as
\begin{equation}
M=\frac{-2\mathcal{C}c_{2}}{3c_{1}}\left( 2k+m^{2}\mathcal{C}%
^{2}c_{2}\right) .
\end{equation}

The above equation imposes some constraints on the mass of black holes. In
other words, in order to have a positive mass of black holes, we obtain some
conditions reported in Table. \ref{tab1}. According to the reported
conditions of Table \ref{tab1}, we obtain the acceptable ranges for $c_{1}$,
$c_{2}$ and $k$.

\begin{table}[h]
\caption{Acceptable ranges for $c_{1}$ and $c_{2}$.}
\label{tab1}\centering
\begin{tabular}{||c|c|c|c||}
\hline\hline
$case$ & $c_{1}$ & $c_{2}$ & $k$ \\ \hline\hline
$I$ & $c_{1}>0~~~\ $ & $~\frac{-2}{m^{2}\mathcal{C}^{2}}<c_{2}<0$ & $k=1$ \\
\hline
$II$ & $c_{1}>0~~~\ $ & $0<c_{2}<\frac{2}{m^{2}\mathcal{C}^{2}}$ & $k=-1$ \\
\hline
$III$ & $c_{1}<0~~~\ $ & $c_{2}<0$ & $k=0$ \\ \hline
$IV$ & $c_{1}<0~~~\ $ & $c_{2}<\frac{-2}{m^{2}\mathcal{C}^{2}}$ & $k=1$ \\
\hline
$V$ & $c_{1}<0~~~\ $ & $c_{2}>0$ & $k=1$ \\ \hline
$VI$ & $c_{1}<0~~~\ $ & $c_{2}>0$ & $k=0$ \\ \hline
$VII$ & $c_{1}<0~~~\ $ & $c_{2}>\frac{2}{m^{2}\mathcal{C}^{2}}$ & $k=-1$ \\
\hline\hline
\end{tabular}%
\end{table}


On the other hand, in order to study white dwarfs and neutron stars with
spherical symmetric assumption, we have to consider $k=1$. In addition, our
studies on white dwarfs \cite{white} and neutron stars \cite{Neutron} with
spherical symmetric ($k=1$) in massive gravity showed that the sign of $%
c_{2} $\ should be negative, but the sign of $c_{1}$\ can be both
positive and negative. In order to explain compact objects such as
neutron stars and white dwarfs with spherical symmetric in massive
gravity, the acceptable range for the $c_{1}$ and $c_{2}$ are I
and IV cases in Table \ref{tab1}.

As one can see, the parameters of the conformal gravity in eq. (\ref{para}),
depend on massive graviton and parameters of massive gravity. Therefore, we
may think that gravitons in the conformal gravity can behave like a massive
particle.

\section{Closing remarks}

In this paper, in order to find a correspondence between
black-hole solutions in conformal and massive theories of gravity,
we have analyzed their metric functions in four dimensions.
Comparing these black holes, we have found a correspondence
between them which imposed some constraints on the parameters of
these theories. ndeed, we have shown that these black holes are
the same, when we impose some conditions between the parameters of
conformal and massive gravities as, i)
$s_{0}=k+m^{2}\mathcal{C}^{2}c_{2}$, ii)
$s_{1}=\frac{m^{2}\mathcal{C}c_{1}}{2}$, and iii) $s_{2}=-M$.

On the other hand, in order to have a positive mass of black holes
in massive gravity, we have a obtained some conditions originated
from the correspondence between black holes in conformal and
massive gravities. For more investigations, we have compared the
reported constraints of white dwarfs and neutron stars in massive
gravity with these conditions. We have found seven physical ranges
for studying black holes in massive gravity with different
topologies (cases I to VII in Table \ref{tab1}), and two physical
ranges for evaluating the compact objects such as white dwarfs and
neutron stars in massive gravity with spherical symmetric
assumption. In other words, the suitable ranges were; i) $c_{1}>0
$ \& $\frac{-2}{m^{2}
\mathcal{C}^{2}}<c_{2}<0$, and ii) $c_{1}<0$ \& $c_{2}<\frac{-2}{m^{2}%
\mathcal{C}^{2}}$ (cases I and IV in Table \ref{tab1}).

As future works, one can study the possible relation between the
mentioned theories with electrically charged and higher
dimensional modifications. More fundamentally, it is interesting
to look for a possible relation between the action and the field
equation of conformal gravity and massive theory. In addition,
regardless of a massless spin-2 graviton which is a standard mode
of gravity, one can find that there are additional degrees of
freedom related to massive or conformal theories of gravity. It is
shown that in conformal gravity there is an additional spin-2 mode
which can be a tachyonic or a massive ghost
\cite{Julve,Alvarez-Gaume}. Besides, as regards massive gravity it
is proven that different classes of massive gravity may have
additional degrees of freedom \cite{dRGTII}. It will be
interesting to investigate the degrees of freedom of the mentioned
theory and look for a possible relation between them.

\acknowledgments We thank an anonymous referee for useful comments. We wish
to thank the Shiraz University Research Council. This work has been
supported financially by Research Institute for Astronomy and Astrophysics
of Maragha (RIAAM), Iran.




\begin{thebibliography}{99}
\bibitem{EnglertTG} F. Englert, C. Truffin, and R. Gastmans, Nucl. Phys. B
\textbf{117}, 407 (1976)

\bibitem{Narlikar} J. V. Narlikar, and A. K. Kembhavi, Lett. Nuovo Cim.
\textbf{19}, 517 (1977)

\bibitem{Riegert} R. J. Riegert, Phys. Rev. Lett. \textbf{53}, 315 (1984).

\bibitem{Maldacena} J. Maldacena, [arXiv:1105.5632].

\bibitem{LuP} H. Lu, and C. N. Pope, Phys. Rev. Lett. \textbf{106}, 181302
(2011).

\bibitem{Bars} I. Bars, P. Steinhardt, and N. Turok, Phys. Rev. D \textbf{89}%
, 043515 (2014).

\bibitem{AnastasiouO} G. Anastasiou, and R. Olea, Phys. Rev. D \textbf{94},
086008 (2016).

\bibitem{MannheimK} P. D. Mannheim, and D. Kazanas, Astrophys. J. \textbf{342%
}, 635 (1989).

\bibitem{Mannheim2006} P. D. Mannheim, Prog. Part. Nucl. Phys. \textbf{56},
340 (2006).

\bibitem{MannheimO} P. D. Mannheim, and J. G. O'Brien, Phys. Rev. Lett.
\textbf{106}, 121101(2011).

\bibitem{Mannheim2012} P. D. Mannheim, Found. Phys. \textbf{42}, 388 (2012).

\bibitem{BergshoeffRd} E. Bergshoeff, M. De Roo, and B. De Wit, Nucl. Phys.
B \textbf{182}, 173 (1981).

\bibitem{deWit} B. De Wit, J. W. van Holten, and A. van Proeyen, Nucl. Phys.
B \textbf{184}, 77 (1981).

\bibitem{Adler} S. L. Adler, Rev. Mod. Phys. \textbf{54}, 729 (1982).

\bibitem{Hooft} G. 't Hooft, Found. Phys. \textbf{41}, 1829 (2011).

\bibitem{BambiMR} C. Bambi, L. Modesto, and L. Rachwal, JCAP. \textbf{05},
003 (2017).

\bibitem{ChakrabartyBBM} H. Chakrabarty, C. A. Benavides-Gallego, C. Bambi,
and L. Modesto, JHEP. \textbf{03}, 013 (2018).

\bibitem{Zhouetal} M. Zhou, et al., Phys. Rev. D \textbf{98}, 024007 (2018).

\bibitem{BerkovitsW} N. Berkovits, and E. Witten, JHEP. \textbf{08}, 009
(2004).

\bibitem{LiuT} H. Liu, and A. A. Tseytlin, Nucl. Phys. B \textbf{533}, 88
(1998).

\bibitem{Balasubramanian} V. Balasubramanian, E. Gimon, D. Minic, and J.
Rahmfeld, Phys. Rev. D \textbf{63}, 104009 (2001).

\bibitem{FP} M. Fierz, and W. Pauli, Proc. R. Soc. Lond. A \textbf{173}, 211
(1939).

\bibitem{BD} D. G. Boulware, and S. Desser, Phys. Lett. B \textbf{40}, 227
(1972).

\bibitem{dRG} C. de Rham, and G. Gabadadze, Phys. Rev. D \textbf{82}, 044020
(2010).

\bibitem{dRGT} C. de Rham, G. Gabadadze, and A. J. Tolley, Phys. Rev. Lett.
\textbf{106}, 231101 (2011).

\bibitem{dRGTI} K. Hinterbichler, Rev. Mod. Phys. \textbf{84}, 671 (2012).

\bibitem{dRGTII} C. de Rham, Living Rev. Relativ. \textbf{17}, 7 (2014).

\bibitem{Kareeso} P. Kareeso, P. Burikham, and T. Harko, Eur. Phys. J. C
\textbf{78}, 941 (2018).

\bibitem{Yamazaki} M. Yamazaki, T. Katsuragawa, S. D. Odintsov, and S.
Nojiri, [arXiv:1812.10239].

\bibitem{BHMassI} T. M. Nieuwenhuizen, Phys. Rev. D \textbf{84}, 024038
(2011).

\bibitem{BHMassII} H. Kodama, and I. Arraut, Prog. Theor. Exp. Phys. \textbf{%
2014}, 023E02 (2014).

\bibitem{BHMassIII} J. Xu, L. -M. Cao, and Y. -P. Hu, Phys. Rev. D \textbf{91%
},124033 (2015).

\bibitem{BHMassIV} S. G. Ghosh, L. Tannukij, and P. Wongjun, Eur. Phys. J. C
\textbf{76}, 119 (2016).

\bibitem{BHMassV} P. Li, X. -z. Li, and P. Xi, Phys. Rev. D \textbf{93},
064040 (2016).

\bibitem{BHMassVI} P. Li, X. -z. Li, and X. -h. Zhai, Phys. Rev. D \textbf{94%
}, 124022 (2016).

\bibitem{BHMassVII} D. -C. Zou, R. Yue, and M. Zhang, Eur. Phys. J. C
\textbf{77}, 256 (2017).

\bibitem{BHMassVIII} S. Chougule, S. Dey, B. Pourhassan, and M. Faizal, Eur.
Phys. J. C \textbf{78}, 685 (2018).

\bibitem{BHMassIX} S. Fernando, Mod. Phys. Lett. A \textbf{33}, 1850177
(2018).

\bibitem{BHMassX} S. -Z. Yang, Q. -C. Ding, and Z. -W. Feng,
[arXiv:1810.05645].

\bibitem{BHMassXI} B. Liu, Z. -Y. Yang, and R. -H. Yue, [arXiv:1810.07885].

\bibitem{Vegh} D. Vegh, [arXiv:1301.0537].

\bibitem{ZhangL} H. Zhang, and X. Z. Li, Phys. Rev. D \textbf{93}, 124039
(2016).

\bibitem{DvaliGS} G. Dvali, G. Gabadadze, and M. Shifman, Phys. Rev. D
\textbf{67}, 044020 (2003).

\bibitem{DvaliHK} G. Dvali, S. Hofmann, and J. Khoury, Phys. Rev. D \textbf{%
76}, 084006 (2007).

\bibitem{DeffayetDG} C. Deffayet, G. Dvali, and G. Gabadadze, Phys. Rev. D
\textbf{65}, 044023 (2002).

\bibitem{GumrukcuogluLM} A. E. Gumrukcuoglu, C. Lin, and S. Mukohyama, JCAP.
\textbf{11}, 030 (2011).

\bibitem{Gratia} P. Gratia, W. Hu, and M. Wyman, Phys. Rev. D \textbf{86},
061504 (2012).

\bibitem{Will} C. M. Will, Living Rev. Relativ. \textbf{17}, 4 (2014).

\bibitem{Mohseni} M. Mohseni, Phys. Rev. D \textbf{84}, 064026 (2011).

\bibitem{GumrukcuogluI} A. E. Gumrukcuoglu, S. Kuroyanagi, C. Lin, S.
Mukohyama, and N. Tanahashi, Class. Quantum Grav. \textbf{29}, 235026 (2012).

\bibitem{white} B. Eslam Panah, and H. L. Liu, Phys. Rev. D \textbf{99}, 104074 (2019).

\bibitem{Neutron} S. H. Hendi, G. H. Bordbar, B. Eslam Panah, and S.
Panahiyan, JCAP. \textbf{07}, 004 (2017).

\bibitem{vanmass} S. H. Hendi, R. B. Mann, S. Panahiyan, and B. Eslam Panah,
Phys. Rev. D \textbf{95}, 021501(R) (2017).

\bibitem{heat} S. H. Hendi, B. Eslam Panah, S. Panahiyan, H. Liu, and X. -H.
Meng, Phys. Lett. B \textbf{781}, 40 (2018).

\bibitem{remnant} B. Eslam Panah, S. H. Hendi, and Y. C. Ong,
[arXiv:1808.07829].

\bibitem{Bergshoeff} E. A. Bergshoeff, O. Hohm, J. Rosseel, and P. K.
Townsend, Phys. Rev. D \textbf{83}, 104038 (2001).

\bibitem{Alishahiha} M. Alishahiha, and R. Fareghbal, Phys. Rev. D \textbf{83%
}, 084052 (2001).

\bibitem{Gullu} I. Gullu, M. Gurses, T. C. Sisman, and B. Tekin, Phys. Rev.
D \textbf{83}, 084015 (2011).

\bibitem{Klemm} D. Klemm, Class. Quant. Grav. \textbf{15}, 3195 (1998).

\bibitem{Mannheim1991} P. D. Mannheim, and D. Kazanas, Phys. Rev. D \textbf{%
44}, 417 (1991).

\bibitem{multiH} S. H. Hendi, S. Panahiyan, B. Eslam Panah, and M. Momennia,
Ann. Phys. \textbf{528}, 819 (2016).

\bibitem{LiuPJ} H. Lu, Y. Pang, C. N. Pope, and J. F. Vazquez-Poritz, Phys.
Rev. D \textbf{86}, 044011 (2012).

\bibitem{XuZhao} W. Xu, and L. Zhao, Phys. Lett. B \textbf{736}, 214 (2014).

\bibitem{XuSZ} H. Xu, Y. Sun, and L. Zhao, Int. J. Mod. Phys. D \textbf{26},
1750151 (2017).

\bibitem{Cai2015} R. -G. Cai, Y. -P. Hu, Q. -Y. Pan, and Y. -L. Zhang, Phys.
Rev. D \textbf{91}, 024032 (2015).

\bibitem{HendiEP} S. H. Hendi, B. Eslam Panah, and S. Panahiyan, JHEP.
\textbf{11}, 157 (2015).

\bibitem{Julve} J. Julve, and M. Tonin, Nuovo Cim. B \textbf{46}, 137 (1978).

\bibitem{Alvarez-Gaume} L. Alvarez-Gaume, A. Kehagias, C. Kounnas, D. Lust,
and A. Riotto, Fortschritte der Physik. \textbf{64}, 176 (2016).
\end{thebibliography}
\end{document}